\documentclass[12 pt]{article}
\usepackage{amssymb,amsthm,amsmath,color,graphics}
\theoremstyle{plain}
\newtheorem{theorem}{Theorem}
\newtheorem{mytheorem}{Theorem}
\newtheorem{lemma}[theorem]{Lemma}
\newtheorem{corollary}[theorem]{Corollary}

\newtheorem{fact}[theorem]{Fact}
\newtheorem{remark}[theorem]{Remark}
\newtheorem{conjecture}[theorem]{Conjecture}
\theoremstyle{definition}
\newtheorem{definition}{Definition}

\newtheorem{example}{Example}

\newcommand{\myket}[1]{|#1\rangle}

\newcommand{\myketbra}[2]{|#1\rangle\langle #2 |}

\newcommand{\ket}[1]{\ensuremath{\left|#1\right>}}
\newcommand{\bra}[1]{\ensuremath{\left<#1\right|}}
\newcommand{\braket}[2]{\ensuremath{\left<#1|#2\right>}}

\newcommand{\ketbra}[2]{| #1 \rangle\langle #2 |}

\newcommand{\be}{\begin{equation}}
\newcommand{\ee}{\end{equation}}
\newcommand{\bc}{\begin{center}}
\newcommand{\ec}{\end{center}}
\newcommand{\bea}{\begin{eqnarray}}
\newcommand{\eea}{\end{eqnarray}}

\renewcommand{\span}{\textrm{span}}

\def\opone{\leavevmode\hbox{\small1\kern-3.8pt\normalsize1}}

\newcommand{\0}{\mathbf{0}}
\newcommand{\1}{\mathbf{1}}
\newcommand{\ketz}{\ket{\0}}
\newcommand{\keto}{\ket{\1}}
\newcommand{\braz}{\bra{\0}}
\newcommand{\brao}{\bra{\1}}
\newcommand{\wid}[1]{{\lbrace #1 \rbrace}}

\def\({\left(}
\def\){\right)}

\begin{document}

\bc { \bf UNIVERSAL QUANTUM COMPUTATION IN A HIDDEN BASIS 
}

\medskip
\renewcommand{\thefootnote}{\fnsymbol{footnote}}
{\footnotesize L. M. Ioannou$^{1,}$\footnote{lmi@iqc.ca} and M.
Mosca$^{1,2,}$\footnote{mmosca@iqc.ca} \\ {\it
$^1$Institute for Quantum Computing, University of Waterloo, \\
200 University Avenue, Waterloo, Ontario, N2L 3G1, Canada \\
$^2$Perimeter Institute for Theoretical Physics\\31 Caroline Street
North, Waterloo, Ontario, N2L 2Y5, Canada}} \ec

\begin{quote}{\small
Let $\ketz$ and $\keto$ be two states that are promised to come from
known subsets of orthogonal subspaces, but are otherwise unknown.
Our paper probes the question of what can be achieved with respect
to the basis $\{\ketz,\keto\}^{\otimes n}$ of $n$ logical qubits,
given only a few copies of the unknown states $\ketz$ and $\keto$. A
phase-invariant operator is one that is unchanged under the relative
phase-shift $\keto \mapsto e^{i \theta}\keto$, for any $\theta$, of
all of the $n$ qubits.  We show that phase-invariant unitary
operators
 can be implemented exactly with no copies and that phase-invariant
states
 can be prepared exactly with at most $n$ copies each of $\ket{\0}$
and $\ket{\1}$; we give an explicit algorithm for state preparation
that is efficient for some classes of states (e.g. symmetric
states).  We conjecture that certain non-phase-invariant operations
are impossible to perform accurately without many copies. Motivated
by optical implementations of quantum computers, we define ``quantum
computation in a hidden basis'' to mean executing a quantum
algorithm with respect to the phase-shifted hidden basis $\{\ketz,
e^{i\theta}\keto\}$, for some potentially unknown $\theta$; we give
an efficient approximation algorithm for this task, for which we
introduce an analogue of a coherent state of light, which serves as
a bounded quantum phase reference frame encoding $\theta$.  Our
motivation was quantum-public-key cryptography, however the
techniques are general. We apply our results to quantum-public-key
authentication protocols, by showing that a natural class of digital
signature schemes for classical messages is insecure.  We also give
a protocol for identification that uses many of the ideas discussed
and whose security relates to our conjecture (but we do not know if
it is secure).}
\end{quote}

\section{Introduction}\label{sec_intro}

We consider a new quantum-information-theoretic problem; let us
first define the problem and then summarize our results and their
significance.

Suppose $\mathcal{S}$ is a $d$-dimensional complex vector space with
computational basis $B = \{\ket{i}: i = 0,1,\ldots, d-1\}$.  Assume
that we have the ability to do universal quantum computation (with
respect to $B$) in $\mathcal{S}$.  Let $\mathcal{S}_0 = \span(B_0)$
and $\mathcal{S}_1 = \span(B_1)$ be two orthogonal subspaces of
$\mathcal{S}$ such that
\begin{eqnarray}
\mathcal{S} = \mathcal{S}_0 \oplus \mathcal{S}_1
\end{eqnarray}
and $B = B_0 \cup B_1$, where the union of the orthonormal bases
$B_0$ and $B_1$ is disjoint.  Assume that $B_0$ and $B_1$ are known,
so that we can perform universal quantum computation with respect to
each of them. For all $b\in \{0,1\}$, let $A_b$ be a set of pure
state vectors,
\begin{eqnarray}
A_b \subset \mathcal{S}_b,
\end{eqnarray}
whose classical description is known, such that no two elements in
$A_b$ are equal up to global phase.

\begin{definition}[Hidden basis]
Let $\ketz$ be a state in $A_0$ and let $\keto$ be a state in $A_1$,
where $A_0$ and $A_1$ are defined above.
 These states define a \emph{hidden (computational) basis}
$\{\ketz,\keto\}^{\otimes n}$ of $n$ logical qubits.  We call this a
``hidden basis'' since in general the choices of $\ketz$ and $\keto$
will not be known.
\end{definition}

\begin{remark}[Notation]
We use boldfaced ket-labels to denote the elements of a hidden
basis.
\end{remark}

\noindent Assuming the ability to do universal quantum computation
in $\mathcal{S}^n := \mathcal{S}^{\otimes n}$ (with respect to the
computational basis $B^{\otimes n}$ of $n$ $d$-dimensional qudits),
we investigate the number of copies of $\ket{\0}$ and $\ket{\1}$
that are required to perform unitary operations and to prepare
quantum states defined with respect to the hidden basis
$\{\ketz,\keto\}^{\otimes n}$. Note that the question is well
defined by virtue of the known classical descriptions of $A_0$ and
$A_1$, which disambiguate the global phases of the states $\ketz$
and $\keto$.

In Section \ref{sec_theta_known}, we show that any phase-invariant
(see Definition \ref{def_phinv}) unitary operator on our $n$ logical
qubits is exactly implementable without requiring any copies of the
states $\ket{\0}$ and $\ket{\1}$.  We then show (see Theorem
\ref{prop2}) that any phase-invariant density operator on $n$
logical qubits is exactly preparable from at most $n$ copies each of
$\ket{\0}$ and $\ket{\1}$.  We then give an explicit, efficient
algorithm for creating symmetric states, based on Ref. \cite{KM01},
that easily generalizes to creating any phase-invariant state.  For
non-phase-invariant unitary operators, such as the logical Hadamard
gate, we conjecture that a large number of copies of $\ket{\0}$ and
$\ket{\1}$ is needed; we give a precise conjecture, in a simplified
framework, in the Appendix.  Our conjecture adds to the important
discussion of what can and cannot be done in quantum mechanics.
Knowing the limitations of a physical (computational) theory is
intrinsically interesting, but no-go theorems can also be used as
building blocks for other useful results. As an example of how one
might use this conjecture for a new kind of cryptographic protocol,
in Section  \ref{sec_app} we present a cryptographic protocol for
identification and explain how our conjecture relates to its
security.

Note that, in practice, performing a quantum computation in the
basis $\{\myket{0}, \myket{1}\}^{\otimes n}$, for the known
qubit-states $\myket{0}$ and $\myket{1}$, is actually equivalent to
performing a computation in the phase-shifted basis $\{\myket{0},
e^{i \phi} \myket{1} \}^{\otimes n}$, where one replaces $\myket{1}$
with $e^{i \phi} \myket{1}$ in all the operations.  For example, in
optical implementations, one typically assumes that a laser outputs
coherent states $\sum_{w=0}^{\infty}(\gamma^w/\sqrt{w!})\myket{w}$
with a random, unknown, but consistent, phase parameter $\phi =
\textrm{arg}(\gamma)$ (see, e.g., Ref. \cite{BRST06}). When using
these coherent states to drive transformations in the qubits, this
is equivalent to performing the entire computation in the basis
$\{\myket{0}, e^{i \phi} \myket{1} \}^{\otimes n}$. It is essential
that the experimentalist maintains a consistent phase reference for
the duration of the computation, but the actual value of $\phi$ is
unimportant. Thus, in Section \ref{sec_theta_unknown}, we consider
the problem of quantum computing with respect to the phase-shifted
hidden basis $\{\ketz,e^{i\theta}\keto\}^{\otimes n}$ for
potentially unknown $\theta \in [0,2\pi)$.  We show (see Theorem
\ref{thm_UniQCHB}) that it is possible to approximate universal
quantum computation in the phase-shifted hidden basis
$\{\ketz,e^{i\theta}\keto\}^{\otimes n}$, given a small phase
reference state encoding $\theta$ (see Definition
\ref{def_phaserefState}) that is analogous to a coherent light
state. It follows (see Corollary \ref{cor_UQCHB}) that, by using a
small number of copies of $\ketz$ and $\keto$, one can prepare such
a phase reference state for unknown and uniformly random $\theta$
and thus carry out approximate universal computation in the
phase-shifted hidden basis $\{\ketz,e^{i\theta}\keto\}^{\otimes n}$
for unknown and uniformly random $\theta$, in analogy to the optical
implementation described above.

Our motivation for considering computing in a hidden basis is rooted
in quantum-public-key cryptography, a framework, introduced in Ref.
\cite{qphGC01}, in which the public keys are copies of a particular
quantum state encoding a classical private key.\footnote{Note that
the number of copies in public circulation must be limited, so that
an adversary at the very least cannot take all the copies, measure
them, and get a sufficiently good estimate of the private key.}~ The
goal of this type of cryptography is to achieve the best of both the
quantum and classical worlds: the information-theoretic security of
several known quantum cryptographic protocols (e.g. quantum key
distribution \cite{May97,SP00} and symmetric-key
message-authentication \cite{BCGST02}) and the advantages (over
symmetric-key cryptography) of a modern public-key infrastructure
(see e.g. Ref. \cite{MvOV96} for details). Unfortunately, it has
been shown in Ref. \cite{BCGST02} that it is generally impossible to
sign arbitrary quantum states, which means that such a quantum
public-key infrastructure may be difficult (if not impossible) to
attain.  Nevertheless, it is important to determine to what extent
quantum-public-key cryptography is feasible.

Our focus in this paper is on authentication schemes, where the
owner, Alice, of the private key attempts to prove to another party,
based on the assumption that this party has an authentic copy of
Alice's public key, that it is indeed Alice who constructed a
certain message (in the case of a digital signature scheme) or
participated in a particular interaction\footnote{Such an
interaction is assumed not to be susceptible to a man-in-the-middle
attack \cite{MvOV96}.}~ (in the case of an identification scheme).
The only known secure quantum-public-key signature scheme is the
one-time digital signature scheme for classical messages
in Ref. \cite{qphGC01}
; the scheme is one-time in that it can only be used to sign one
message securely before the public keys need to be refreshed.  A
natural next step is thus to find a ``reusable'' quantum digital
signature scheme for classical messages or prove none exists.
In the context of authentication schemes, we define \emph{reusable}
to mean that Alice can use the same private key to sign many
different messages or prove her identity many times, but a fresh
copy of the public key is needed for each verification instance.
In Section \ref{sec_app}, we describe a rather natural and general
cryptographic framework, based on hidden bases, that may be suitable
for reusable quantum-public-key authentication schemes.  We show
(see Corollary \ref{cor_nogo}) how our abovementioned state
preparation result is a cryptanalytic tool, rendering insecure a
class of quantum digital signature schemes within the framework,
thus effectively extending the original no-go theorem for quantum
digital signatures in Ref. \cite{BCGST02}. Finally, in an attempt to
stimulate further research in reusable quantum-public-key
authentication schemes, we give a protocol for identification that
uses many of the ideas discussed (but we do not know if the protocol
is secure).

\section{Phase-invariant operators}\label{sec_theta_known}

Let $\mathcal{H}$ denote the span of $\{\ketz, \keto\}$; thus,
$\mathcal{H}^{n} := \mathcal{H}^{\otimes n}$ denotes the span of the
hidden basis for $n$ logical qubits:
\begin{eqnarray}
\mathcal{H}^{n} = \span(\{\ketz, \keto \}^{\otimes n}).
\end{eqnarray}
For any bit-string $y=y_1y_2\cdots y_n \in \{0,1\}^n$, let
\begin{eqnarray}
\ket{\mathbf{y}}:=\ket{\mathbf{y_1}}\ket{\mathbf{y_2}}\cdots\ket{\mathbf{y_n}}.
\end{eqnarray}

For any $\theta \in [0,2\pi]$, let $U(\theta)$ be the
\emph{phase-shift by $\theta$ (with respect to the basis
$\{\ket{\mathbf{y}}: y \in \{0,1\}^n\}$) operator} on
$\mathcal{H}^{n}$ such that
\begin{eqnarray}
U(\theta): \ket{\mathbf{y}} \mapsto  e^{i
H(y)\theta}\ket{\mathbf{y}},
\end{eqnarray}
where $H(y):= \sum_j y_j$ is the Hamming weight of $y$.

\begin{definition}[Phase invariant]\label{def_phinv}
Let $T$ be any operator on $\mathcal{H}^{n}$.  Then $T$ is
\emph{phase(-shift) invariant (with respect to $\{\ket{\mathbf{y}}:
y \in \{0,1\}^n\}$)} if and only if
\begin{eqnarray}\label{eq_DefPhaseInv}
U(\theta) T U(\theta)^\dagger = T
\end{eqnarray}
for all $\theta \in [0,2\pi]$.
\end{definition}

\noindent Define the \emph{weight} of $\ket{\mathbf{y}}$ to be
$H(y)$, and define the weight-$w$ subspace of $\mathcal{H}^{ n}$ as
\begin{eqnarray}
\mathcal{H}^{n}_w := \textrm{span}(\{\ket{\mathbf{x}} \in
\mathcal{H}^{ n}: H(x)=w\}).
\end{eqnarray}

\begin{fact}\label{Fact_BlockDiag} A linear operator $T$ on $\mathcal{H}^{ n}$ is phase invariant if and
only if it is block diagonal with respect to the decomposition
$\mathcal{H}^{n} =\oplus_{w=0}^n \mathcal{H}^{n}_w$.
\end{fact}
\proof{Writing $T = \sum_{y,z} T_{y,z}
\ketbra{\mathbf{y}}{\mathbf{z}}$ and $U(\theta) = \sum_x e^{i H(x)
\theta} \ketbra{\mathbf{x}}{\mathbf{x}}$, it is easy to show that
\begin{eqnarray}
U(\theta)TU(\theta)^\dagger = \sum_{y,z} e^{i \theta (H(y) - H(z))}
T_{y,z} \ketbra{\mathbf{y}}{\mathbf{z}}
\end{eqnarray}
and thus Eq. (\ref{eq_DefPhaseInv}) may be rewritten
\begin{eqnarray}\label{eq_PhaseInvEquiv}
T_{y,z} = e^{i \theta (H(y) - H(z))} T_{y,z}\textrm{,      for all }
y,z \in \{0,1\}^n.
\end{eqnarray}
If $T$ is block diagonal, then the equality in Eq.
(\ref{eq_PhaseInvEquiv}) holds for all $\theta$ when $H(y)\neq H(z)$
because $T_{y,z}=0$; this equality always holds for all $\theta$
when $H(y) = H(z)$.  To prove the other direction, note that, if
$H(y)\neq H(z)$ and Eq. (\ref{eq_PhaseInvEquiv}) holds for all
$\theta$, then $T_{y,z}$ must be zero (otherwise one could divide
both sides by $T_{y,z}$ and get a contradiction for some value of
$\theta$).}

\subsection{Exact implementation/preparation of phase-invariant unitary/density operators}

The following lemma implies that, despite our limited knowledge
about $\ketz$ and $\keto$, we can exactly implement any
phase-invariant unitary operator $V$ on $\mathcal{H}^{n}$, given its
matrix (explicitly) with respect to the hidden basis. The lemma
guarantees that we can find a matrix representation of $V$ with
respect to the computational basis $B^{\otimes n}$ of
$\mathcal{S}^{n}$, and then use this to effect $V$ on
$\mathcal{H}^{n}$; this implementation of $V$ is algorithmically
exact (though, in practice, error correction would likely need to be
used; we assume perfect quantum channels throughout this paper).

\begin{lemma}\label{prop1} Given the matrix representation of
a phase-invariant unitary operator $V$ on $\mathcal{H}^{n}$ with
respect to the hidden basis $\{\myket{\0}, \myket{\1}\}^{\otimes
n}$, one can compute the matrix representation of an operator $V'$
on $\mathcal{S}^{n}$ with respect to the computational basis
$B^{\otimes n}$, such that $V'\left.\right|_{\mathcal{H}^{ n}} = V$.
\end{lemma}
\proof{
Since $V$ is phase invariant, Fact \ref{Fact_BlockDiag} implies it
is block diagonal with respect to $\oplus_{w=0}^n \mathcal{H}^{
n}_w$, and can thus be written $V = \oplus_{w=0}^n V_w$, for $V_w$
unitary on $\mathcal{H}^{ n}_w$.  Let $\{\ket{w,z}:z=1,2,\ldots,{n
\choose w}\}$ be the natural ordered basis for $\mathcal{H}^{ n}_w$,
that is, $\ket{w,z} \in \{\myket{\0}, \myket{\1}\}^{\otimes n}$ for
all $(w,z)$. The operator $V_w$ is specified by ${n \choose w}$
equations of the form
\begin{eqnarray}\label{eqn_hiddenmap}
\ket{w,z}\mapsto \sum_{k=1}^{{n \choose w}} c^w_{k,z} \ket{w,k},
\end{eqnarray}
which we can read off the given matrix for $V$.  If, for each
$z\in\{1,2,\ldots,{n \choose w}\}$, we fix $\pi^w_z$ to be any
permutation on $n$ objects\footnote{Here, the $n$ objects will be
the $n$ components of a vector (that functions as the label for a
ket).  A binary string is considered a vector of zeros and ones.}~
such that
\begin{eqnarray}
\ket{\pi^w_z(\mathbf{0}^{n-w}\mathbf{1}^w)} = \ket{w,z},
\end{eqnarray}
we can rewrite Eq. (\ref{eqn_hiddenmap}) as
\begin{eqnarray}\label{eqn_hiddenmappi}
\ket{\pi^w_z(\mathbf{0}^{n-w}\mathbf{1}^w))}\mapsto \sum_{k=1}^{{n
\choose w}} c^w_{k,z} \ket{\pi^w_k (\mathbf{0}^{n-w}\mathbf{1}^w)}.
\end{eqnarray}
Let $S^{ n}_w$ denote the ``weight-$w$'' subspace of $S^{ n}$:
\begin{eqnarray}
S^{ n}_w := \span \lbrace \ket{ \pi^w_z (c)   } \in S^{ n}: z \in
\{1,2,\ldots,{n \choose w}\}, c \in (B_0)^{n-w}\times
(B_1)^w\rbrace.
\end{eqnarray}
It suffices to show how to compute the matrix representation with
respect to $B^{\otimes n}$ of a unitary operator $V'_w$ on $S^{
n}_w$ such that $V'_w\left.\right|_{\mathcal{H}^{ n}_w} = V_w$, for
each $w$; thus, fix $w$.

Assuming $d_0:=\dim\mathcal{S}_0$ and $d_1:=\dim\mathcal{S}_1$, we
can substitute into Eq. (\ref{eqn_hiddenmappi}) the two equations
\begin{eqnarray}
\ketz := \sum_{i=1}^{d_0} \alpha_i \ket{a_i},\hspace{5mm} \keto :=
\sum_{j=1}^{d_1} \beta_j \ket{b_j},
\end{eqnarray}
where $B_0=\{\ket{a_i}\}_i$ and $B_1 =\{\ket{b_j}\}_j$, and get,
after changing the order of summations,
\begin{eqnarray}
\sum_{i_1,\ldots,i_{n-w}}\sum_{j_1,\ldots,j_{w}}
\alpha_{i_1}\cdots\alpha_{i_{n-w}}\beta_{j_1}\cdots\beta_{j_{w}}\ket{\pi^w_z(a_{i_1},\ldots,a_{i_{n-w}},b_{j_1},\ldots,b_{j_{w}})}\hspace{10mm}\\
\mapsto\sum_{i_1,\ldots,i_{n-w}}\sum_{j_1,\ldots,j_{w}}
\alpha_{i_1}\cdots\alpha_{i_{n-w}}\beta_{j_1}\cdots\beta_{j_{w}}
\left( \sum_{k=1}^{{n \choose w}} c^w_{k,z} \ket{\pi^w_k
(a_{i_1},\ldots,a_{i_{n-w}},b_{j_1},\ldots,b_{j_{w}})}
 \right).
\end{eqnarray}
 Consider the mapping defined by the ${n \choose w}d_0^{n-w}d_1^w$
equations of the form
\begin{eqnarray}\label{eqns_Vprimew}
\ket{\pi^w_z(a_{i_1},\ldots,a_{i_{n-w}},b_{j_1},\ldots,b_{j_{w}})}
\mapsto\sum_{k=1}^{{n \choose w}} c^w_{k,z} \ket{\pi^w_k
(a_{i_1},\ldots, a_{i_{n-w}},b_{j_1},\ldots,b_{j_{w}})}
\end{eqnarray}
for all $\ket{a_{i_l}}\in \{\ket{a_i}\}_i$ (for all
$l=1,2,\ldots,n-w$), for all $\myket{b_{j_{l'}}}\in \{\ket{b_j}\}_j$
(for all $l'=1,2,\ldots,w$), and for all $z = 1,2,\ldots {n \choose
w}$. We claim that this mapping well-defines a suitable $V'_w$.
 Indeed, it is easy to see that the $d_0^{n-w}d_1^w$ subspaces (indexed
 by $({i_1},\ldots,{i_{n-w}},{j_1},\ldots,{j_{w}})$)
\begin{eqnarray}
\span\lbrace\myket{\pi^w_z(a_{i_1},\ldots,a_{i_{n-w}},b_{j_1},\ldots,b_{j_{w}})}:
z \in \{1,2,\ldots,{n \choose w}\}\rbrace
\end{eqnarray}
are mutually orthogonal, and that the mapping is unitary on each of
these subspaces by unitarity of $V_w$. Since $\dim S^{ n}_w = {n
\choose w}d_0^{n-w}d_1^w$, the mapping well-defines a unitary
operator on $S^{ n}_w$.  The matrix entries for $V'_w$ can be read
off of Eqs (\ref{eqns_Vprimew}).}

Lemma \ref{prop1} allows us to prove the following theorem, which
implies that, despite our limited knowledge of $\ketz$ and $\keto$,
we can prepare a copy of any phase-invariant density operator on
$\mathcal{H}^{ n}$, given its matrix with respect to the hidden
basis.

\begin{mytheorem}[Exact preparation of phase-invariant states]\label{prop2}
Given the matrix representation of a phase-invariant density
operator $\rho$ on $\mathcal{H}^{ n}$ with respect to the hidden
basis $\{\myket{\0}, \myket{\1}\}^{\otimes n}$, one can prepare a
copy of $\rho$ using at most $n$ copies each of $\ket{\0}$ and
$\ket{\1}$.
\end{mytheorem}
\proof{Lemma \ref{prop1} implies that, in order to prepare any
phase-invariant pure state $\ket{\phi} \in \mathcal{H}^{ n}_w$, it
suffices to have $(n-w)$ copies of $\ket{\0}$ and $w$ copies of
$\ket{\1}$. To see this, note that there exists a unitary operator
$U_w$ on $\mathcal{H}^{ n}_w$ mapping
\begin{eqnarray}
\myket{\0}^{\otimes (n-w)}\myket{\1}^{\otimes w} \mapsto \ket{\phi},
\end{eqnarray}
and that $U_w$ is (trivially) phase invariant.  Since $\rho$ is just
a probabilistic distribution of phase-invariant pure states (because
it is block diagonal), it follows that $\rho$ is preparable using at
most $n$ copies each of $\ket{\0}$ and $\ket{\1}$ (assuming one can
sample from this probability distribution).}

\subsection{Algorithm for exact state preparation of phase-invariant states}\label{sec_ExactStatePrepKM01}

Theorem \ref{prop2} does not address the question of efficiency.
Indeed, in some cases, the required unitary operation (denoted $U_w$
in the proof) is efficient, as demonstrated by the following
example.

Let $\myket{S^n_w}$ be the normalized symmetric sum of all $n
\choose w$ states in $\{\ketz,\keto\}^{\otimes n}$ that have weight
$w$:
\begin{eqnarray}
\myket{S^n_w} := \frac{1}{\sqrt{n \choose w}} \sum_{x\in \{0,1\}^n:
H(x)=w} \ket{\mathbf{x}}.
\end{eqnarray}
As we now explain, the algorithm for state generation in Ref.
\cite{KM01} can be adapted to transform \begin{eqnarray}
\myket{\0}^{\otimes (n-w)}\myket{\1}^{\otimes w}\mapsto
\myket{S^n_w}.\end{eqnarray} Fix $n$ and assume $0 < w < n$.
Hypothetically, suppose we had a copy of $\myket{S^n_w}$ and we
measured the registers one-by-one from the left in the basis
$\{\ketz,\keto\}$. Denote the binary outcome of measuring a register
by 0 (if the register was in state $\myket{\0}$) or 1 (if the
register was in state $\myket{\1}$). Let $X_i$, for
$i=1,2,\ldots,n$, denote the random variable representing the
outcome of measuring register $i$ (registers are enumerated from
left to right).  For $j=2,3,\ldots, n$ and for any $x\in
\{0,1\}^{j-1}$ and $x_j \in \{0,1\}$, define the probabilities
\begin{eqnarray}
p_{x} &:=& P(X_1\cdots X_{j-1} = x)\\
p_{x_j|x} &:=& P(X_{j} = x_j| X_1\cdots X_{j-1}=x),
\end{eqnarray}
where adjacent bit-values denote string-concatenation and we note
that the definition of $p_{x}$ holds also for $j=n+1$. Then we have
$p_1 = w/n=1-p_0$ and
\begin{eqnarray}
p_{x_j|x}&=&p_{xx_{j}}/p_{x} \\
p_{1|x}&=& (w-H(x))/(l-j+1)\\
&=& 1 -p_{0|x}.
\end{eqnarray}
Define the shorthand notation, for nonnegative integers $d$ and
$c\geq d$,
\begin{eqnarray}\label{def_numberstate}
\myket{d^\wid{c}}:=\myket{\0}^{\otimes (c-d)}\myket{\1}^{\otimes d}.
\end{eqnarray}
We can prepare $\myket{S^n_w}$ by starting with $\myket{w^\wid{n}}$,
and then applying a sequence $U_1, U_2,\ldots, U_n$ of phase
invariant unitary operators. The first operator will be
\begin{eqnarray}
U_1:\hspace{2mm}\myket{w^\wid{n}} \mapsto \sqrt{p_0}\myket{\0}
\myket{w^\wid{n-1}}+ \sqrt{p_1}\myket{\1}\myket{(w-1)^\wid{n-1}}.
\end{eqnarray}
For each $j=2,3,\ldots,n$ and for any $x=x_1x_2\cdots x_{j-1}\in
\{0,1\}^{j-1}$, we define the
 operators, for any $d \in \{1,2,\ldots,r \}$,
\begin{eqnarray}
U_j:\hspace{2mm}\myket{\mathbf{x}}\myket{d^\wid{n-(j-1)}} \mapsto
\myket{\mathbf{x}}\left(
\sqrt{p_{0|x}}\myket{\0}\myket{d^\wid{n-j}}+
\sqrt{p_{1|x}}\myket{\1}\myket{(d-1)^\wid{n-j}}\right).
\end{eqnarray}
Each $U_j$ performs an operation similar to a
root-\textsc{swap} operator\footnote{
For $0<\alpha<1$, we can define a \emph{root-\textsc{swap}-like}
operator, which maps $\myket{\0}\myket{\0}\mapsto
\myket{\0}\myket{\0}$, $\myket{\0}\myket{\1}\mapsto
\alpha\myket{\0}\myket{\1} +i\sqrt{1-\alpha^2}\myket{\1}\myket{\0}$,
$\myket{\1}\myket{\0}\mapsto i\sqrt{1-\alpha^2}\myket{\0}\myket{\1}
+\alpha\myket{\1}\myket{\0}$, and $\myket{\1}\myket{\1}\mapsto
\myket{\1}\myket{\1}$. When $\alpha=1/\sqrt{2}$, this is the
root-\textsc{swap} operator.}, controlled on registers 1 through
$(j-1)$, swapping register $j$ with the next closest register to the
right whose state is orthogonal to the subspace containing
$\myket{\0}$. Our $U_j$ also has built into it a final
phase-clean-up operation, controlled on registers $1$ through $j$,
which removes the imaginary factor of $i$ arising from the
root-\textsc{swap} operation.  We can now show that 
$\myket{S^n_w}=U_n\cdots U_2U_1\myket{w^\wid{n}}$.
A straightforward induction (similar to that in Ref. \cite{KM01})
shows that, after $U_j$ is applied, the state of the $n$ registers
is
\begin{eqnarray}\label{line_IndAssump}
\sum_{x \in \{0,1\}^j} \sqrt{p_x} \myket{\mathbf{x}}
\myket{(w-H(x))^\wid{n-j}}
\end{eqnarray}
so that, after $U_{n}$ is applied, the state is $\sum_{x\in
\{0,1\}^n}\sqrt{p_x}\myket{\mathbf{x}}=\myket{S^n_w}$.

The above algorithm for creating $\myket{S^n_w}$ can be generalized
to create any $\myket{\eta} \in \mathcal{H}^{ n}_w$ (and hence any
phase-invariant density operator): it is clear that it can be
generalized to create any state in $\mathcal{H}^{ n}_w$ that has
real coefficients; we refer to Ref. \cite{KM01} for how to create
the correct phases of any complex coefficients of $\myket{\eta}$
efficiently. The general algorithm is efficient as long as the
conditional probabilities $p_{x_j|x}$ are efficiently computable, as
in the case of the symmetric states $\myket{S^n_w}$.

\begin{example}
Consider the state $\rho = \int_0^{2
\pi}\ketbra{\varphi(\theta)}{\varphi(\theta)} d\theta$, where
\begin{eqnarray}
\ket{\varphi(\theta)}=\left(\frac{\myket{\0} + e^{i
\theta}\myket{\1}}{\sqrt{2}}\right)^{\otimes n} =
\frac{1}{2^{n/2}}\sum_{w=0}^n \sqrt{n \choose w} e^{i w\theta
}\myket{S^n_w}.
\end{eqnarray}
We see that $\rho$ is phase invariant and equal to $\sum_{w=0}^n {n
\choose w}2^{-n} \myketbra{S^n_w}{S^n_w}$.  Theorem \ref{prop2}
implies we can prepare $\rho$ using at most $n$ copies each of
$\ket{\0}$ and $\ket{\1}$, and the above algorithm implies we can do
so efficiently.
\end{example}

\subsection{Beyond phase invariance?}\label{sec_BeyondPhaseInv}

One may of course ask about unitary operators and quantum states
that are not phase invariant.  Regarding the former, recall the
finite universal set $\{H,S,T,c\textrm{-\emph{Z}}\}$ of gates
\cite{NC00}, where
\begin{eqnarray}\label{line_UniversalSet}
H = \frac{1}{\sqrt{2}}\left[
  \begin{array}{ c c }
     1 & 1\\
     1 & -1
  \end{array} \right],\hspace{2mm}
S = \left[
  \begin{array}{ c c }
     1 & 0\\
     0 & i
  \end{array} \right],\hspace{2mm}
T = \left[
  \begin{array}{ c c }
     1 & 0\\
     0 & e^{i\pi/4}
  \end{array} \right],\hspace{2mm}
\mbox{c-}Z = \left[
  \begin{array}{ c c c c}
     1 & 0 & 0 & 0\\
     0 & 1 & 0 & 0\\
     0 & 0 & 1 & 0\\
     0 & 0 & 0 & -1\\
  \end{array} \right],
\end{eqnarray}
and the matrices are defined with respect to $\{\ketz,\keto\}$ (in
the case of the one-qubit gates) and $\{\ketz,\keto\}^{\otimes 2}$
(in the case of the $c\textrm{-\emph{Z}}$ gate).  Since $S$, $T$,
and $c\textrm{-\emph{Z}}$ are phase invariant, we see that it is the
presumed inability to implement the Hadamard gate $H$ exactly that
prevents us from performing exact universal quantum computation in
the hidden basis $\{\ketz,\keto\}^{\otimes n}$. Indeed, we
conjecture that, for worst-case $A_0$ and $A_1$ (recall their
definitions in the Introduction), a large number of copies of
$\ket{\0} \in A_0$ and $\ket{\1} \in A_1$ are necessary to implement
one Hadamard gate or to prepare one copy of $(\ket{\0} +
\ket{\1})/\sqrt{2}$, if the implementation/preparation is well
approximated for every $\ketz \in A_0$ and every $\keto \in A_1$;
see Remark \ref{rem_sq} and the Appendix for a precise conjecture in
a simplified framework.

\section{Approximate universal computation in a hidden basis}\label{sec_theta_unknown}

Recall our discussion in the Introduction, where we noted that
quantum computing with respect to the hidden basis $\{\ketz,
\keto\}^{\otimes n}$ is equivalent to computing with respect to the
phase-shifted hidden basis $\{\ketz, e^{i \theta}\keto\}^{\otimes
n}$, as long as $\theta$ is consistent throughout the entire
computation.

\begin{definition}[Quantum computation in a hidden
basis]\label{def_UniQCHB}~ Let $\rho_0$ be a density operator on
$\mathcal{H}^{ n}$ and let $W$ be a unitary operator on
$\mathcal{H}^{ n}$.  To carry out the \emph{quantum computation (of
$(\rho_0, W)$) in the (phase-shifted) hidden basis} $\{\ketz, e^{i
\theta}\keto\}^{\otimes n}$ means to effect the operation
\begin{eqnarray}\label{eqn_universalqcinahiddenbasis}
\rho_0 \mapsto \rho_0':= U(\theta)WU(\theta)^\dagger \rho_0
U(\theta) W^\dagger U(\theta)^\dagger,
\end{eqnarray}
given a copy of $\rho_0$ and a classical description of $W$. We say
the computation is carried out \emph{approximately, with fidelity
$\sqrt{1-\epsilon^2}$}, if we effect an operation $\rho_0 \mapsto
\rho_0''$ such that $\rho_0''$ has fidelity $\sqrt{1-\epsilon^2}$
with $\rho_0'$.
\end{definition}

\begin{remark}[Compatibility of phase references]\label{rem_compatiblephase}
In Definition \ref{def_UniQCHB}, note that $\rho_0'$ will be
equivalent to $W \rho_0 W^\dagger$ up to conjugation by $U(\theta)$
if $\rho_0$ is phase invariant.   More generally, if $\rho_0 =
U(\theta)\sigma_0 U(\theta)^\dagger$ for some $\theta$-independent
$\sigma_0$ on  $\mathcal{H}^n$, then $\rho_0'$ will be equivalent to
$W \sigma_0 W^\dagger$ up to conjugation by $U(\theta)$.  The latter
condition (while including all phase-invariant $\rho_0$, in which
case $\rho_0 = \sigma_0$) includes any non-phase-invariant $\rho_0$
that is (somehow) already defined with respect to the phase-angle
$\theta$, e.g., $\rho_0 = ((\ketz + e^{i \theta}\keto)(\braz +
e^{-i\theta}\brao)/2)^{\otimes n}$ (corresponding to $\sigma_0 =
((\ketz + \keto)(\braz + \brao)/2)^{\otimes n}$).
\end{remark}

\noindent Note that we have not unnecessarily complicated Definition
\ref{def_UniQCHB} by including a notion of measurement: any
measurement can be expressed as an extra unitary operation (on a
possibly larger space) plus a projective measurement in the hidden
computational basis $\{\ketz,\keto\}^{\otimes n}$; the extra unitary
operation may be absorbed into $W$ and the projective measurement
easily simulated by measuring with respect to the computational
basis of $\mathcal{S}^{ n}$.

We now show how we can achieve approximate universal quantum
computation in the hidden basis $\{\ketz, e^{i
\theta}\keto\}^{\otimes n}$ efficiently, given a phase reference
state encoding $\theta$ that is used in an analogous manner to how
coherent light states are used to drive qubit transformations in an
optical implementation of a quantum computer.

Recalling the discussion in Section \ref{sec_BeyondPhaseInv}, we
note that the set $C_\theta:=\{H_\theta, S, T, \textrm{$c$-$Z$}\}$
is universal for quantum computation with respect to the
phase-shifted hidden basis $\{\ketz, e^{i \theta} \keto\}^{\otimes
n}$, where the phase-shifted Hadamard gate is defined as
\begin{eqnarray}
H_\theta :=   \frac{1}{\sqrt{2}}\left[
  \begin{array}{ c c }
     1 & e^{-i \theta}\\
     e^{i \theta} & -1
  \end{array} \right],
\end{eqnarray}
where the matrix is with respect to the hidden basis $\{\ketz,
\keto\}$. Since the other gates in $C_\theta$ are phase invariant,
it thus suffices that we show how to implement (approximately) the
gate $H_\theta$ many times, each time on an arbitrary input, and
each time with respect to the same value of unknown and uniformly
random $\theta \in [0, 2 \pi)$.  We can actually implement, for any
$\alpha \in [0,1]$, the generalized phase-shifted Hadamard gate
\begin{eqnarray}\label{hadamard1}
H_\theta(\alpha):\hspace{5mm}\myket{\0} &\mapsto& \alpha\myket{\0} +
\sqrt{1-\alpha^2} e^{i \theta}\myket{\1}\\\label{hadamard2} e^{i
\theta}\myket{\1} &\mapsto& \sqrt{1-\alpha^2}\myket{\0} - \alpha
e^{i \theta}\myket{\1}.
\end{eqnarray}
Just as before, we will make use of phase-invariant
root-\textsc{swap}-like operations, which introduce imaginary $i$
factors; thus, it will be more convenient to directly implement the
gate,
\begin{eqnarray}
{G_\theta(\alpha)}:\hspace{2mm} \myket{\0} &\mapsto& \alpha\myket{\0} + \sqrt{1-\alpha^2}ie^{i \theta}\myket{\1}\\
\myket{\1} &\mapsto& \sqrt{1-\alpha^2}ie^{-i \theta}\myket{\0}
+\alpha\myket{\1},
\end{eqnarray}
and then we have, for example, $Z S {G_\theta(\alpha)}S Z =
H_\theta(\alpha)$, where $Z$ is the (phase-invariant) Pauli-$Z$ gate
\begin{eqnarray}
Z = \left[
  \begin{array}{ c c }
     1 & 0\\
     0 & -1
  \end{array} \right]
\end{eqnarray}
and $S$ is defined in Eq. (\ref{line_UniversalSet}).  Also, for
clarity of exposition, we will assume $\alpha=1/\sqrt{2}$ (but we
will indicate in footnotes how the procedure is modified for general
$\alpha$).  Let
\begin{eqnarray}
G_\theta := G_\theta (1/\sqrt{2}).
\end{eqnarray}

Consider how one might effect the gate $G_\theta$, given some phase
reference state that encodes $\theta$ and presumably depends on
$\ketz$ and $\keto$. What form could such a state take?
Inspired by a coherent light state,
$\sum_{w=0}^{\infty}(\gamma^w/\sqrt{w!})\myket{w}$, 
we make the following definition.
\begin{definition}[Phase reference state]\label{def_phaserefState}
For any $\theta \in [0,2\pi)$ and positive integer $t$, a
\emph{phase reference state} (having \emph{size} $t$ and
\emph{encoding} $\theta$) is
\begin{eqnarray}\label{eqn_def_of_Psi_k}
\myket{\Psi^t_\theta} &:=& \sum_{w=1}^{t} e^{i w\theta}
\myket{w^\wid{t}},
\end{eqnarray}
where $\myket{w^\wid{t}}$ is defined in Eq. (\ref{def_numberstate}).
\end{definition}
\noindent Each occurrence of $\myket{\1}$ (respectively,
$\myket{\0}$) in the state $\myket{w^\wid{t}}$ is analogous to one
photon (respectively, one vacuum).  Thus, it will be convenient to
refer to the state $\myket{w^\wid{t}}$ as a \emph{$\1$-number
state}, in analogy with the photon-number state
$\myket{w}$.\footnote{In the right-hand side of Eq.
(\ref{eqn_def_of_Psi_k}), we could have had the summation start at
$w=0$; we choose $w=1$ to make the analysis that follows cleaner,
while not significantly affecting the quality of the approximation.}

\begin{remark}[Freedom in definition of phase reference state]\label{rem_qc}
Note that since we are not restricted to the standard interaction
Hamiltonians present in Nature (because we are just mimicking such
interactions using a universal quantum computer), our version of the
coherent states, as well as how they interact with the other
systems, looks slightly different; we need only mimic some of the
main properties of coherent light states, for example, that the
argument (phase angle) of successive coefficients scales linearly
with the photon number. In general, to achieve an efficient
approximation, there is some freedom in the choice of the moduli of
the coefficients of our phase reference states.  For clarity of
presentation, we have chosen to use the simplest coefficients, with
constant modulus. However, one could specify a cost function and
optimize the moduli accordingly, with modest gain in the quality of
the approximation.\footnote{States that are similar to our
coherent-state analogues $\myket{\Psi^t_{\theta}}$ have
independently been used in Ref. \cite{qphLTW}.
 In their application, the phase parameter is not an issue. Rather,
they use this property of coherent states behaving more and more
``classically'', i.e. with less and less disturbance to the coherent
state, as the size $t$ of the coherent state gets larger. In their
case, a larger coherent state corresponds to a larger amount of
shared entanglement, and they used this to show that more
entanglement always improves the success probability of their
protocol. Generally, all such states are forms of ``embezzling''
states \cite{vDH03,qphvDH02}.}
\end{remark}

Define the phase-invariant controlled-root-\textsc{swap} gate $U$,
which, for fixed $t$, acts on $(t+1)$ equally-sized registers, and
in particular operates as follows:
\begin{eqnarray}\label{eqn_U1}
U: \myket{\0}\myket{a^\wid{t}}&\mapsto& \myket{\0}\myket{a^\wid{t}}
+ {i}\myket{\1}\myket{(a-1)^\wid{t}}\\\label{eqn_U2}
\myket{\1}\myket{b^\wid{t}}&\mapsto&
i\myket{\0}\myket{(b+1)^\wid{t}} + \myket{\1}\myket{b^\wid{t}},
\end{eqnarray}
for all $a = 1,2,\ldots,t$ and all $b =
0,1,\ldots,(t-1)$.\footnote{Here is a complete description of $U$:
If the registers are enumerated $0,1,\ldots,t$ starting from the
left, then $U$ applies the root-\textsc{swap} to registers $0$ and
$j$ (for $j\geq 1$) exactly when \emph{either} registers $0$ through
$(j-1)$ are in the state subspace containing $\myket{\0}$ and
registers $j,\ldots,t$ are in the subspace containing $\myket{\1}$
\emph{or} registers 1 through $j$ are in the subspace containing
$\myket{\0}$ and registers 0 and $j+1,\ldots, t$ are in the subspace
containing $\myket{\1}$ (and, otherwise, $U$ acts as the identity
operator). In the case of general $\alpha$, use the
root-\textsc{swap}-like operator (defined in a previous footnote)
instead of the root-\textsc{swap}.} ~Let $\myket{\phi}$ be an
arbitrary one-logical-qubit pure state to which we want to apply
${G_{\theta}}$. Applying $U$ to
$\myket{\phi}\myket{\Psi_\theta^{t}}$, the resulting state contains
the term
\begin{eqnarray}
\sqrt{\frac{t-2}{t}}({G_\theta}
\myket{\phi})\left(\frac{1}{\sqrt{t-2}}\sum_{w=2}^{t-1}e^{ i \theta
w} \myket{w^\wid{t}}\right).
\end{eqnarray}
The other (``junk'') terms are of $\1$-number $0$, $1$, and $t$.
Note that terms of different $\1$-number are orthogonal.
Recall that it suffices to compute the minimum gate fidelity over
all \emph{pure} inputs $\myket{\phi}$, because of the joint
concavity of the fidelity \cite{NC00}.  Thus, the ${G_\theta}$ gate
was effectively applied with minimum
fidelity $\sqrt{(t-2)/t}$, 
and the rightmost $(t+1)$ registers have this same fidelity with the
state $\myket{\Psi_{\theta}^{t}(1)}$, where we define
\begin{eqnarray}
\myket{\Psi_{\theta}^t(i')}:=
(1/\sqrt{t-2i'})\sum_{w=1+i'}^{t-i'}e^{i \theta w}\myket{w^\wid{t}},
\end{eqnarray}
for $i'\geq 0$. To implement a second ${G_\theta}$ gate, we use
these \emph{same} rightmost $t$ registers (whose state is close to
$\myket{\Psi_{\theta}^{t}(1)}$).  And so on. In general, we find
that $U\myket{\phi}\myket{\Psi^t_\theta(i')}$ equals
\begin{eqnarray}\label{term}
\sqrt{\frac{t-2(i'+1)}{t-2i'}}({G_\theta}\myket{\phi})\myket{\Psi^t_{\theta}(i'+1)}
\end{eqnarray}
plus orthogonal ``junk'', where $\myket{\phi}$ is, again, an
arbitrary pure state of a logical qubit. The gate fidelity of the
$l$th approximation of ${G_\theta}$ is thus at least
$\sqrt{(t-2l)/t}$ (for sufficiently large $t$).  Thus, we have the
following theorem:

\begin{mytheorem}[Approximate universal quantum computation in a hidden basis]\label{thm_UniQCHB}
Let $\rho_0$ be a density operator on $\mathcal{H}^{ n}$ and suppose
$W$ is a unitary operator on $\mathcal{H}^{ n}$ that can be
decomposed into phase-invariant gates and at most $l$ Hadamard
gates.  Given $\epsilon >0$ and a copy of the reference state
$\myket{\Psi_\theta^t}$ (defined above), one can carry out the
quantum computation in the hidden basis $\{\ketz, e^{i
\theta}\keto\}^{\otimes n}$ of $(\rho_0, W)$ approximately, with
fidelity $\sqrt{1 - \epsilon^2}$, if $t \geq {\lceil} 2l/\epsilon^2
{\rceil}$.
\end{mytheorem}


\noindent 
We say that the approximation algorithm in Theorem \ref{thm_UniQCHB}
is efficient because the size of the phase reference state need only
scale linearly with the number of Hadamard gates implemented, for
constant $\epsilon$.  In Section \ref{sec_Security}, we show how
Theorem \ref{thm_UniQCHB} can be applied by an adversary to mount a
weak attack on the cryptographic protocol we present in Section
\ref{sec_idscheme}.

We end this section with the following corollary, which summarizes
how one can use copies of $\ketz$ and $\keto$ to create a phase
reference state $\ket{\Psi^t_\theta}$, for unknown and uniformly
random $\theta$, in order to carry out approximate universal quantum
computation in a hidden basis.

\begin{corollary}[Computation in a hidden basis for unknown and random $\theta$]\label{cor_UQCHB}
Let $\rho_0$ be a density operator on $\mathcal{H}^{ n}$ and suppose
$W$ is a unitary operator on $\mathcal{H}^{ n}$ that can be
decomposed into phase-invariant gates and at most $l$ Hadamard
gates.  Given $\epsilon >0$ and $t$ copies each of $\ketz$ and
$\keto$, one can carry out the quantum computation in the hidden
basis $\{\ketz, e^{i \theta}\keto\}^{\otimes n}$ of $(\rho_0, W)$
\emph{for unknown and uniformly random $\theta \in [0, 2\pi)$},
i.e., effect the operation
\begin{eqnarray}
\rho_0 \mapsto \rho_0':= \frac{1}{2\pi}\int_\theta
U(\theta)WU(\theta)^\dagger \rho_0 U(\theta) W^\dagger
U(\theta)^\dagger d\theta,
\end{eqnarray}
approximately, with fidelity $\sqrt{1 - \epsilon^2}$, if $t \geq
{\lceil} 2l/\epsilon^2 {\rceil}$.
\end{corollary}
\proof{Noting that
\begin{eqnarray}
\frac{1}{2\pi}\int_{0}^{2 \pi} d\theta
\myketbra{\Psi^{t}_\theta}{\Psi^{t}_\theta} =
\frac{1}{t}\sum_{w=1}^t \myketbra{w^\wid{t}}{w^\wid{t}},
\end{eqnarray}
we can thus prepare the state $\myket{\Psi^{t}_\theta}$ for a
uniformly random value of $\theta$ by preparing a uniformly random
$\1$-number state, which is easy to do with $t$ copies each of
$\ketz$ and $\keto$.  The statement then follows from Theorem
\ref{thm_UniQCHB}.}

\noindent  We say that the approximation algorithm in Corollary
\ref{cor_UQCHB} is efficient because the number of copies of $\ketz$
and $\keto$ need only scale linearly with the number of Hadamard
gates implemented, for constant $\epsilon$.  Note that if $\rho_0$
is phase invariant and we are interested in measuring (after
executing $W$) a phase-invariant Hermitian observable $M$, then we
can apply Corollary \ref{cor_UQCHB} to carry out approximate quantum
computation in a hidden basis and effectively measure the observable
$M$ on the state $W \rho_0 W^\dagger$.\footnote{We have specified
that the observable $M$ be phase invariant for two reasons: (1) so
that the measurement statistics from measuring $M$ on $\rho_0'$ are
the same as if one measured $M$ on $W\rho_0 W^\dagger$; and (2) so
that we can implement the measurement with no approximation error,
as follows. Each block $M_w$ of any phase-invariant observable
$M=\oplus_{w=0}^n M_w$ can be diagonalized by a phase-invariant
unitary $V_w$.  Thus, we can implement $M$ by first implementing
$\oplus_{w=0}^n V_w^\dagger$ and then measuring in the hidden
computational basis $\{\ketz, \keto\}^{\otimes n}$.}~ We hope that
Corollary \ref{cor_UQCHB} finds application (perhaps in conjunction
with Theorem \ref{thm_UniQCHB}) in interactive protocols (e.g.
interactive proofs), where the parties variously create/send/receive
and perform quantum operations on input states ($\rho_0$), output
states ($\rho_0'$), and phase reference states.

\section{Applications to public-key authentication}\label{sec_app}

Our motivation is information-theoretically-secure
\emph{quantum-public-key} cryptography, a framework for which was
first proposed by Gottesman and Chuang in Ref. \cite{qphGC01}; we
describe an example of that framework now.

Let $A_{\delta} \subset \mathbf{C}^M$ be a set of superpolynomially
(in ${\log}(M)$) many quantum states such that, for every distinct
$\myket{\psi}$ and $\myket{\phi}$ in $A_{\delta}$,
\begin{eqnarray}\label{pwdist}
|\braket{\psi}{\phi}| \leq \delta
\end{eqnarray}
for some positive constant $\delta <1$.  The states in $A_{\delta}$
are sometimes called \emph{quantum fingerprints}, and explicit
constructions for such $A_{\delta}$, with $|A_{\delta}| \in
2^{\Omega(M)}$, are known \cite{BCWW01}.  A (succinct) classical
description of $A_{\delta}$ is published.

As part of the key generation procedure, Alice randomly chooses a
$\myket{\psi}$ uniformly from $A_{\delta}$ and keeps this choice
secret, but she authentically distributes (e.g., by trusted courier)
a limited number of copies of $\myket{\psi}$ among several members
of the public (including Bob). The classical description of
$\myket{\psi}$, which can be encoded by a bit-string of length
$\Theta({\log}(|A_{\delta}|))$, serves as part of the \emph{private
key}, while several authentic copies of the state $\myket{\psi}$
serve as part of the \emph{public key}. Assuming each copy of
$\myket{\psi}$ is an $O({\log}(M))$-qubit state, the maximum number
of bits of information one can extract from $T$ copies of
$\myket{\psi}$ is $O(T{\log}(M))$, by the Holevo bound \cite{Hol73,
NC00}.  Thus, as long as $T \ll {\log}(|A_{\delta}|)/{\log}(M)$, the
part of the private key corresponding to $\ket{\psi}$ is protected
even if $T$ copies of $\myket{\psi}$ exist.

The full private/public key would typically consist of several
independent instances of this setup, that is, Alice independently
chooses several states $\ket{\psi} \in A_{\delta}$ and distributes
the corresponding copies.  This naturally allows for protocols that
are the composition of independent instances of an atomic protocol,
or \emph{kernel}, that succeeds with only a certain probability.
Repeating the kernel sufficiently many times, with independent
$\ket{\psi}$-values each time, can amplify the success probability
to an acceptable level. In the case of authentication schemes, ``to
succeed'' means ``to correctly `accept' or `reject' a purported
message or entity''.

A more general framework can be obtained by allowing the public key
to consist of additional systems that may depend on the private key;
we will use this more general framework in Section
\ref{sec_idscheme}.

We are primarily interested in public-key \emph{authentication}
protocols --- either for classical messages, as in digital signature
schemes, or for entities, as in identification schemes --- a general
approach to which is the following.  Suppose that $A_{\delta}$ also
satisfies the condition
\begin{eqnarray}\label{nozerocomp}
 \braket{0}{\psi}=0
\end{eqnarray}
for all $\myket{\psi} \in A_{\delta}$ and some known $\myket{0} \in
\mathbf{C}^M$. Alice can easily create states like $\myket{0} +
\myket{\psi}$ (we will sometimes omit normalization factors) and,
more generally, she can perform any computation in the basis
$\{\ket{0},\ket{\psi}\}^{\otimes n}$.

\begin{remark}[No-Squashing Conjecture]\label{rem_sq} In general, it is not known how to perform
such computations efficiently, and we conjecture that it takes
superpolynomially (in ${\log}(M)$) many copies of a uniformly random
$\myket{\psi} \in A_{\delta}$ to prepare even one copy of
$\myket{0}+ \myket{\psi}$, if the procedure is to work for all
$\myket{\psi}\in A_{\delta}$.  We call this the \emph{No-Squashing
Conjecture}.  We call the task of creating $\myket{0}+ \myket{\psi}$
for every $\myket{\psi} \in A_{\delta}$, given copies of
$\myket{\psi}$, \emph{squashing}. See the Appendix for more details
about squashing and the No-Squashing Conjecture.
\end{remark}

\noindent The hope is to use this framework, for example, for Alice
to convince Bob that she has prepared some state (like a signature)
that no one without full knowledge of $\myket{\psi}$ could have done
with only the limited number of copies of the state $\myket{\psi}$
available. Moreover, we are interested in \emph{reusable} schemes,
by which we mean that the same private key (and corresponding public
keys) can be used to identify Alice many times or sign many messages
from Alice (recall that the digital signature scheme in Ref.
\cite{qphGC01} is not reusable: it can only be used to sign one
message).  This framework may be suitable for reusable schemes,
because, as suggested by our example protocol in Section
\ref{sec_idscheme}, it seems to allow protocols where Alice does not
divulge (to a verifier or adversary) a significant amount of extra
information about the private key (beyond that which is already
available from all the copies of the public key), yet retains some
advantage over the adversary.

\begin{remark}[Notation]
In our cryptographic setting, the unknown state $\myket{\psi}$ is
now playing the role of $\keto$. Thus, for this section of the
paper, we redefine all the objects (e.g., $\myket{w^\wid{t}}$,
$\myket{\Psi^t_\theta}$, $\myket{S^n_w}$, $H$) that depend on
$\myket{\0}$ and $\myket{\1}$, by replacing each occurrence of
$\myket{\0}$ with $\myket{0}$ and each occurrence of $\myket{\1}$
with $\myket{\psi}$.  
\end{remark}

\subsection{Insecurity of a class of digital signature schemes for classical
messages}\label{sec_nogo}

Theorem \ref{prop2} can be interpreted as a restriction on any
digital signature scheme for classical messages in the above
framework.  Before we state the result, we give a more detailed
description of such a signature scheme.

Suppose that, in the key generation procedure described in the
previous section, Alice chose $K$ independent values $\ket{\psi_k}$
from $A_\delta$, $k=1,2,\ldots, K$.  Let $x$ denote the message to
be signed.  We assume that the full signature state for message $x$
is a $J$-fold tensor product of states
\begin{eqnarray}
\bigotimes_{j=1}^J \sigma_j(\myket{\psi_{k(j,x)}}),
\end{eqnarray}
where $k(j,x) \in \{1,2,\ldots, K\}$ is a publicly known function
depending on the particular scheme, and each
$\sigma_j(\myket{\psi_{k(j,x)}})$ is a density operator on $\span
(\{\ket{0},\myket{\psi_{k(j,x)}}\}^{\otimes n})$ such that the
coefficients of $\sigma_j(\myket{\psi_{k(j,x)}})$ with respect to
the basis $\{\ket{0},\myket{\psi_{k(j,x)}}\}^{\otimes n}$ are
publicly known.\footnote{These coefficients, which may also depend
on the message $x$, are known in that they do not depend on the
private key (classical description of $\myket{\psi_{k(j,x)}}$). This
allows the adversary to compute the conditional probabilities
required for our state preparation algorithm of Section
\ref{sec_ExactStatePrepKM01}.}~ Note that, in general,
$\sigma_j(\ket{\psi})$ need not equal $\sigma_{j'}(\ket{\psi})$ when
$j\neq j'$.

We assume further that the full verification procedure breaks up
into $J$ independent procedures, each denoted $P_j$, one for each
$\sigma_j(\myket{\psi_{k(j,x)}})$. Thus, if no adversary interferes,
Bob would apply the procedure $P_j$ to
$\sigma_j(\myket{\psi_{k(j,x)}})$ (using his copy of the public key
and the message) and obtain some measurement statistics. After doing
this for all $j=1,2,\dots,J$, he would process all the statistics
and determine whether to ``accept'' or ``reject'' the
message-signature pair.

\begin{corollary}[Insecurity of a class of digital signature
schemes]\label{cor_nogo} Suppose there is a signature scheme with
signature state and verification procedure as described above.
Suppose further that an adversary can obtain $n$ copies of
$\myket{\psi_{k(j,x)}}$, for all $j=1,2,\ldots,J$. Then the scheme
is not information-theoretically secure if, for all
$j=1,2,\ldots,J$, the procedure $P_j$ applied to the state
$\sigma_j(e^{i \theta}\myket{\psi_{k(j,x)}})$ produces the same
statistics as if it were applied to $\sigma_j(\ket{\psi_{k(j,x)}})$,
for any $\theta \in [0,2\pi]$.
\end{corollary}
\proof{From Theorem \ref{prop2}, it follows that the adversary can
create the uniform mixture over $\theta \in [0,2\pi)$ of
$\sigma_j(e^{i \theta}\myket{\psi_{k(j,x)}})$, because this mixture
is phase invariant. The procedure $P_j$ applied to this mixture will
also produce the same statistics as if it were applied to
$\sigma_j(\ket{\psi_{k(j,x)}})$, thus the scheme is not
information-theoretically secure.}

We note that, for example, Corollary \ref{cor_nogo} applies to any
scheme such that the only public key states available for use in
verification of $\sigma_j(\myket{\psi_{k(j,x)}})$ are copies of
$\myket{\psi_{k(j,x)}}$ (and no other state dependent on the private
key) and the verification procedure uses the copies of
$\myket{\psi_{k(j,x)}}$ only as input to
\textsc{swap}-tests\footnote{Recall that the
\emph{\textsc{swap}-test} \cite{BCWW01,qphGC01} of two registers
(labelled 2 and 3) in the states $\myket{\xi}_2$ and
$\myket{\phi}_3$ is a measurement (with respect to the computational
basis $\lbrace \myket{0}_1, \myket{1}_1\rbrace$) of the control
register (labelled 1) of the state
\begin{eqnarray}
(H_1 \otimes I_2 \otimes
I_3)(c-\textsc{swap}_{2,3}){(\myket{0}_1+\myket{1}_1)}\myket{\xi}_2\myket{\phi}_3/{\sqrt{2}},
\end{eqnarray}
where $H_1$ is the Hadamard gate (applied to register 1) and
$c-\textsc{swap}_{2,3}$ is the controlled-\textsc{swap} gate.  The
probability that the state is $\ket{0}_1$ immediately after the
measurement --- which corresponds to a \emph{pass} --- is
$(1+|\braket{\xi}{\phi}|^2)/2$. When the registers 2 and 3 are in
the mixed states $\rho$ and $\rho'$, this probability is
$(1+\textrm{tr} (\rho\rho'))/2$.}~ (or similar tests for symmetry
under permutations \cite{BCWW01,NI09}). Generally, the corollary
implies that the verification procedure for any secure signature
scheme in this framework, where the adversary can obtain
sufficiently many copies of $\myket{\psi_{k(j,x)}}$, will have to
exploit the global phase of the state vector $\myket{\psi_{k(j,x)}}
\in A_{\delta}$ determined by its classical description.

\subsection{Example of a cryptographic protocol in this framework}\label{sec_idscheme}

We now give an example of a cryptographic protocol that incorporates
many of the ideas we have presented.  The protocol is actually a
translation of the honest-verifier identification scheme of Ref.
\cite{qphIM09} into our hidden basis setting. The following is an
intuitive description (adapted from Section 4.7.5.1 in Goldreich's
book \cite{Gol01}) of how a secure identification scheme works.

Suppose Alice generates a private key and authentically distributes
copies of the corresponding public key to any potential users of the
scheme, including Bob.  If Alice wants to identify herself to Bob
(i.e. prove that it is she with whom he is communicating), she
invokes the identification protocol by first telling Bob that she is
Alice, so that Bob knows he should use the public key corresponding
to Alice (assuming Bob possesses public keys from many different
people). The ensuing protocol, whatever it is, has the property that
the \emph{prover} Alice can convince the \emph{verifier} Bob (except
perhaps with negligible probability) that she is indeed Alice, but
an adversary Eve cannot fool Bob (except with negligible
probability) into thinking that she is Alice, even after having
listened in on the protocol between Alice and Bob or having
participated as a (devious) verifier in the protocol with Alice
several times. An \emph{honest-verifier identification protocol} is
only intended to be secure under the extra assumption that, whenever
Eve engages the prover Alice in the protocol, Eve follows the
verification protocol as if she were honest. Note that no
identification protocol is secure against a person-in-the-middle
attack, where Eve concurrently acts as a verifier with Alice and as
a prover with Bob.

As part of the key generation procedure for our protocol, we assume
Alice has chosen a random $\myket{\psi} \in A_{\delta}$ and has
distributed at most $r$ copies of the state
\begin{eqnarray}
\myket{\psi} (\myket{0} + \myket{\psi}).
\end{eqnarray}
This state is the public key for one iteration of the kernel of the
protocol. The parameter $r$ is the \emph{reusability} parameter,
dictating the maximum number of secure uses of the scheme for a
particular public key.

The kernel of our interactive protocol is the following three steps,
which form a typical ``challenge-response'' interactive proof.
If the kernel is repeated $s$ times in total, then one copy of the
(full) public key (of which there are still $r$ copies in total)
would be
$\otimes_{i=1}^s \myket{\psi_i} (\myket{0} + \myket{\psi_i})$, where
the $\myket{\psi_i}$ are each independently and uniformly randomly
picked from $A_{\delta}$ by Alice. The parameter $s$ is the
\emph{security parameter}, which is chosen after $r$ is fixed.

\begin{enumerate}
\item Bob uses $\ket{\psi}$ to create the symmetric state $\myket{S_1^2}=\ket{0}\ket{\psi} +
\ket{\psi}\ket{0}$ (as shown in Section
\ref{sec_ExactStatePrepKM01}), and sends the leftmost register of
this state to Alice.

\item On the received register, Alice performs the logical Hadamard gate
$H$ and then measures with respect to an orthogonal basis
$\{\myket{0}, \myket{\psi}, \ldots\}$. If the state of the register
immediately after the measurement is $\myket{0}$, then Alice sends
``0'' to Bob; otherwise, Alice sends ``1''.

\item If Bob receives ``1'', then he applies the $Z$ gate
to the register that he kept (that contained half of the symmetric
state he made in Step 1). Finally, Bob \textsc{swap}-tests this
register with the register containing the authentic copy of
$\myket{0} + \myket{\psi}$ (the \textsc{swap}-test is defined in
Section \ref{sec_nogo}).
\end{enumerate}

\noindent After the kernel is repeated $s$ times, Bob ``accepts'' if
all the \textsc{swap}-tests passed; otherwise, Bob ``rejects''.  As
a final specification for the protocol, we also stipulate that Alice
not engage in the protocol more than $r$ times (when there are $r$
copies of the public key in circulation) for a particular value of
the private key.

Before discussing the potential security of this scheme, we note
that the honest protocol is correct because
\begin{eqnarray}
\myket{S^2_1} = (H^{-1}\ket{0})(\ket{0}+\ket{\psi}) -
(H^{-1}\ket{\psi})Z^{-1}(\ket{0}+\ket{\psi});
\end{eqnarray}
that is, Bob's \textsc{swap}-test always passes when the prover is
honest (assuming perfect quantum channels).

\subsubsection{Discussion of potential security}\label{sec_Security}

Within the hidden basis cryptographic framework, we refer to as
\emph{black-box attacks} those attacks where Eve does not use any
information about the structure of $A_{\delta}$ to help her cheat.
In the following discussion, we restrict our attention to the
honest-verifier setting. The following definition of ``security''
suffices for our discussion:\footnote{As in Ref. \cite{Gol01}, our
definition of ``security'' does not include the completeness of the
protocol, which stipulates that honest Bob should always accept when
the prover is honest (this is easily verified for our protocol). Our
definition does not take into account that there may be many
different honest provers. As well, we consider neither the parallel
nor serial composability of the identification protocol. See Ref.
\cite{Gol01} for more details in the classical case.}
\begin{definition}[Security]
An honest-verifier identification protocol (for honest prover Alice
and honest verifier Bob) is \emph{secure with error $\epsilon$} if
the probability that Bob ``accepts'' when any adversary Eve
participates in the protocol as a prover is less than $\epsilon$
(assuming that, whenever Eve engages Alice in the protocol, Eve
follows the verification protocol honestly).
\end{definition}

One obvious black-box attack that Eve could perform is as follows.
Eve can collect $r':= (r-1)$ copies of $\ket{0} + \ket{\psi}$, which
are in the state
\begin{eqnarray}
\left(\frac{\ket{0} + \ket{\psi}}{\sqrt{2}}\right)^{\otimes r'} =
\frac{1}{\sqrt{2^{r'}}} \sum_{w=0}^{r'} \sqrt{r' \choose
w}\myket{S^{r'}_w}.
\end{eqnarray}
Now, assume Eve performs the inverse of the phase-invariant
operation given in Section \ref{sec_ExactStatePrepKM01}, which maps
$\myket{S_w^{r'}} \mapsto \myket{w^\wid{r'}}$ with no error.  The
state thus becomes
\begin{eqnarray}
\frac{1}{\sqrt{2^{r'}}}\sum_{w=0}^{r'} \sqrt{{r' \choose
w}}\myket{w^\wid{r'}}.
\end{eqnarray}
Note that this state is similar to $\myket{\Psi^{r'}_0}$, but for
the coefficients, which now have non-constant moduli (see Remark
\ref{rem_qc}).  Thus, Eve can use this state as a phase reference in
order to mimic Alice's Hadamard gate.

In the black-box (honest-verifier) setting, given any $\epsilon >0$
and $r$, there indeed exists a value of $s$ (dependent on $r$ and
$\epsilon$) such that the protocol is secure with error $\epsilon$
(assuming perfect quantum channels).  For, in this case, the
protocol reduces to the honest-verifier-secure protocol of Ref.
\cite{qphIM09}.  The security proof follows from the work of
Bartlett et al. \cite{BRST06} on bounded quantum reference frames
and is a formalization of the following intuition. Note that Alice
always causes Bob's \textsc{swap}-test to pass. However, Eve's
information about the correct reference frame is limited to her $r$
samples of it (because Eve cannot extract any further information
from Alice). Since only an infinite number of samples should suffice
for Eve to be able to perform a measurement in the logical Hadamard
basis perfectly (as Alice can), there is always some nonzero
probability that Eve causes Bob's \textsc{swap}-test to fail.  With
sufficiently large $s$, Bob will find such a failure (except with
negligible probability).  It turns out that $s \in \Omega(r
\log(r/\epsilon))$ suffices \cite{qphIM09}.

A security proof would of course need to consider all attacks ---
not just black-box ones.  Note that if squashing required only a
small number of copies of $\myket{\psi}$ (see Remark \ref{rem_sq}),
then Eve could prepare more copies of $\myket{0} + \myket{\psi}$ to
use as a phase reference for her approximate implementation of $H$.
This is one way that the security of our scheme depends on the
assumed difficulty of squashing (e.g., the No-Squashing Conjecture);
however, even if Eve could somehow transform her copies of
$\myket{\psi}$ into one or more copies of $\myket{0} +
\myket{\psi}$, the parameter $s$ could be modestly increased to
account for Eve's extra samples of the reference frame, assuming
there exists an $s$ such that the scheme is secure for $r'$ samples
of $\myket{0} + \myket{\psi}$. The security of the scheme depends
more crucially on the weaker conjecture that it is impossible to
perform a measurement with respect to the logical Hadamard basis
$\{\myket{0} \pm \myket{\psi}\}$ (given the limited number of copies
of $\myket{\psi}$) much more efficiently than with our black-box
reference-frame approach (this conjecture is weaker because if Eve
could carry out the measurement, then she could squash).

\section{Closing Remarks}

By exploiting phase invariance and mimicking properties of coherent
states of light, we have shown how to perform various computational
tasks, defined with respect to a hidden computational basis
$\{\ketz,\keto\}^{\otimes n}$, efficiently in the required number of
copies of $\ket{\0}$ and $\ket{\1}$. We have shown that such tasks,
which were previously not known to be possible, have cryptological
application.

We have identified several open problems, including the squashing
problem and the harder problem of performing measurements with
respect to the hidden Hadamard basis $\{\ket{\0} \pm \ket{\1}\}$.
Another open problem is to investigate to what extent state
preparation and universal computation are possible when the
assumption that $\ket{\0}$ and $\ket{\1}$ come from known orthogonal
subspaces is dropped, that is, when the only promise is that
$\braket{\0}{\1}=0$ with $\ketz, \keto \in A$, where $A$ contains no
two states that are equal up to global phase.

\section*{Acknowledgements}

We thank Scott Aaronson for discussions on the squashing problem and
for naming it.  We thank Daniel Gottesman for discussions on
signature and identity schemes.  We are grateful for the referees'
comments on presentation, organization, and technical points. L. M.
Ioannou was supported by the EPSRC and SCALA; most of this work was
done while he was at the University of Cambridge. M. Mosca was
supported by NSERC, QuantumWorks, MITACS, CIFAR, CRC, ORF, OCE,
Government of Canada, and Ontario-MRI.


\begin{thebibliography}{10}

\bibitem{KM01}
Phillip Kaye and Michele Mosca.
\newblock Quantum networks for generating arbitrary quantum states.
\newblock In {\em International Conference on Quantum Information}, OSA
  Technical Digest Series, page PB28, Washington, D.C., June 2001. Optical
  Society of America.

\bibitem{BRST06}
Stephen~D. Bartlett, Terry Rudolph, Robert~W. Spekkens, and Peter~S.
Turner.
\newblock Degradation of a quantum reference frame.
\newblock {\em New J. Phys.}, 8:58, 2006.

\bibitem{qphGC01}
Daniel Gottesman and Isaac~L. Chuang.
\newblock Quantum digital signatures, 2001.
\newblock arXiv:quant-ph/0105032.

\bibitem{May97}
D.~Mayers.
\newblock Unconditionally secure quantum bit commitment is impossible.
\newblock {\em Phys. Rev. Lett.}, 78(17):3414, 1997.

\bibitem{SP00}
Peter~W. Shor and John Preskill.
\newblock Simple proof of security of the {BB84} quantum key distribution
  protocol.
\newblock {\em Phys. Rev. Lett.}, 85:441--444, 2000.

\bibitem{BCGST02}
Howard Barnum, Claude Cr{\'{e}}peau, Daniel Gottesman, Adam Smith,
and Alain
  Tapp.
\newblock Authentication of quantum messages.
\newblock In IEEE Press, editor, {\em Proc. 43rd Annual IEEE Symposium on the
  Foundations of Computer Science (FOCS '02)}, pages 449--458, 2002.

\bibitem{MvOV96}
A.~J~. Menezes, P.~van Oorschot, and S.~Vanstone.
\newblock {\em Handbook of Applied Cryptography}.
\newblock CRC Press LLC, Boca Raton, 1996.

\bibitem{NC00}
M.~Nielsen and I.~Chuang.
\newblock {\em Quantum Computation and Quantum Information}.
\newblock Cambridge University Press, Cambridge, 2000.

\bibitem{qphLTW}
Debbie Leung, Ben Toner, and John Watrous.
\newblock Coherent state exchange in multi-prover quantum interactive proof
  systems, 2008.
\newblock arXiv:0804.4118.

\bibitem{vDH03}
Wim van Dam and Patrick Hayden.
\newblock Universal entanglement transformations without communication.
\newblock {\em Phys. Rev. A}, 67(6):060302(R), 2003.

\bibitem{qphvDH02}
Wim van Dam and Patrick Hayden.
\newblock Embezzling entangled quantum states, 2002.
\newblock arXiv:quant-ph/0201041.

\bibitem{BCWW01}
Harry Buhrman, Richard Cleve, John Watrous, and Ronald de~Wolf.
\newblock Quantum fingerprinting.
\newblock {\em Phys. Rev. Lett.}, 87:167902, 2001.

\bibitem{Hol73}
A.~S. Holevo.
\newblock Statistical problems in quantum physics.
\newblock In Gisiro Maruyama and Jurii~V. Prokhorov, editors, {\em Proceedings
  of the Second Japan-USSR Symposium on Probability Theory}, volume 330 of {\em
  Lecture Notes in Mathematics}, pages 104--119, Berlin, 1973. Springer-Verlag.

\bibitem{NI09}
Georgios~M. Nikolopoulos and Lawrence~M. Ioannou.
\newblock Deterministic quantum-public-key encryption: Forward search attack
  and randomization.
\newblock {\em Phys. Rev. A}, to appear.

\bibitem{qphIM09}
Lawrence~M. Ioannou and Michele Mosca.
\newblock Public-key cryptography based on bounded quantum reference frames.
\newblock arXiv:0903.5156.

\bibitem{Gol01}
O.~Goldreich.
\newblock {\em Foundations of cryptography ({V}olume I): {B}asic tools}.
\newblock Cambridge University Press, Cambridge, 2001.

\bibitem{Sh06}
A.Y. Shiekh.
\newblock The role of quantum interference in quantum computing.
\newblock {\em International Journal of Theoretical Physics}, 45(9):1646--1648,
  2006.

\bibitem{BHW00}
V.~Bu{\v{z}}ek, M.~Hillery, and F.~Werner.
\newblock Universal-{NOT} gate.
\newblock {\em J. Mod. Opt.}, 47, 2000.

\end{thebibliography}


\section*{Appendix: Squashing and the No-Squashing Conjecture}

The squashing problem can be meaningfully (and nontrivially) defined
for a broad class of sets of states --- not just sets of type
$A_{\delta}$. Let $A$ be a set of pure state vectors in
$\mathbf{C}^M$ such that no two distinct elements in $A$ are equal
up to global phase. (Formally, let $\{A(M)\}_{M=1,2,\ldots}$ be a
family of sets $A(M) \subset \mathbf{C}^{M}$ of complex unit vectors
expressed relative to the standard basis
$\{\myket{0},\myket{1},\myket{M-1}\}$. For clarity of presentation,
we usually omit the ``family'' notation $\{\cdot\}_{M=1,2,\ldots}$.)
Let $\epsilon>0$ be the error tolerance parameter.  The pair
$(A,\epsilon)$ specifies an instance of the general squashing
problem, which is to compute, for every $\myket{\psi} \in A$, a
state $\rho$ such that
\begin{eqnarray}
D(\sigma_{0,\psi}, \rho) \leq \epsilon,
\end{eqnarray}
given copies of $\myket{\psi}$, where $D$ is the trace distance
\cite{NC00},
\begin{eqnarray}
\sigma_{0,\psi} := (\myket{0} + \myket{\psi})(\bra{0}+\bra{\psi}),
\end{eqnarray}
and $\myket{0}$ is some known reference state (which, in general,
need not be orthogonal to all $\myket{\psi} \in A$).  Note that the
classical description of $A$ also specifies a global phase for each
possible state vector $\ket{\psi}$, so the ideal target state
$\myket{0} + \myket{\psi}$ is well defined. The set $A$ should be
\emph{nontrivial}, meaning that it should contain elements that are,
pairwise, not orthogonal (as would be the case when $A =
A_{\delta}$); this is analogous to the problem of quantum cloning,
where it is trivial to clone states that are promised to come from a
prescribed orthogonal set.

Define $t(A,\epsilon)$ to be the smallest $t$ for which there exists
a quantum operation $Q$ such that, for all $\myket{\psi} \in A$,
\begin{eqnarray}\label{bb}
D( Q((\ketbra{\psi}{\psi})^{\otimes t}),\sigma_{0,\psi}) \leq
\epsilon.
\end{eqnarray}
A quantum operation $Q$ that, for some $t$, satisfies Eq. (\ref{bb})
for all $\myket{\psi}\in A$ is called an
$(A,\epsilon)$-\emph{squasher}, because it ``squashes'' part of the
generalized Bloch sphere towards its $\myket{0}$-pole.  We view the
number $t(A,\epsilon)$ as measuring the complexity of the squashing
problem instance $(A,\epsilon)$.  We assume all $\ket{\psi}$ in $A$
are reasonably encoded into qubits and thus take the input size of
the problem to be $\log(M)$.

Our cryptographic motivation leads us to look for sets $A$ such that
$t(A,\epsilon)$ is necessarily large (for sufficiently small
$\epsilon$).  For some instances of the general squashing problem,
exponential lower bounds on $t(A,\epsilon)$ might be relatively
easily obtainable, because it might be the case that, for
subexponential values of $t$, the states $\sigma_{0,\psi}$ and
$\sigma_{0,\phi}$ are further in trace distance (or, equivalently,
have lower fidelity) than $\myket{\psi}^{\otimes t}$ and
$\myket{\phi}^{\otimes t}$.
The following example of this shows that the general squashing
problem subsumes the black-box search problem (this fact has already
been pointed out in Ref. \cite{Sh06}, though the author of that
paper assumes that squashing is an easy task).

\begin{example}\label{eg_BBSearch}
Denote by $F$ the quantum Fourier transform on $\mathbf{C}^M$, so
that
\begin{eqnarray}
F\myket{0} = (1/\sqrt{M})\sum_{i=0}^{M-1} \myket{i}.
\end{eqnarray}
For every $j=0,1,\ldots,M-1$, define the state $\myket{j^*}$ via
\begin{eqnarray}
\braket{i}{j^*} &=& \left\{ \begin{array}{ll}
                +1/\sqrt{M} & \mbox{ if $i=j$ }, \\
                -1/\sqrt{M} & \mbox{ if $i\neq j$     }.
                \end{array}
                \right.
\end{eqnarray}
Let $A_2:= \{F^{\dagger}\myket{j^*}:j=0,1,\ldots,M-1\}$ (note that
$A_2$ does \emph{not} have the property that $\braket{0}{\psi}=0$
for all $\myket{\psi} \in A_2$). Noting that $\myket{j^*}+F\myket{0}
\propto \myket{j}$, we can solve the black-box search problem with a
good $(A_2, \epsilon)$-squasher, where $\epsilon$ is a constant
(i.e., if the solution to the search problem is $j$, then the black
box can be used $t$ times to make $t$ copies of $\myket{j^*}$).
Thus, the well-known $\sqrt{M}$ search lower bound applies to
$t(A_2,\epsilon)$. But this is overkill: the lower bound we get from
the fact that quantum operations cannot increase trace distance is
larger when $\epsilon < 1/2$, as we now show.   If $Q$ is an $(A_2,
\epsilon)$-squasher, then we have (for $i\neq j$) 
\begin{eqnarray}
1-2\epsilon &\leq& D( Q((F^{\dagger}\myketbra{i^*}{i^*}F)^{\otimes
t}) , Q((F^{\dagger}\myketbra{j^*}{j^*}F)^{\otimes t}) )
\\&\leq& D ((\myketbra{i^*}{i^*})^{\otimes t} ,
(\myketbra{j^*}{j^*})^{\otimes t} )
\\&\leq& \sqrt{1- |\braket{i^*}{j^*}|^{2t}}
\\&=&\sqrt{1-(1-4/M)^{2t}},
\end{eqnarray}
 so that $t \geq
{{\log}(4\epsilon(1-\epsilon))}/{2\log(1-4/M)}$ and thus, for
example, $t(A_2,\frac{1}{3},1)\in \Omega(M)$.  The first line
follows from several applications of the triangle inequality and the
fact that the trace distance between the squasher's ideal outputs is
1. In subsequent lines, we have used the well-known relationship
between trace distance and fidelity (see Ref. \cite{NC00}) and the
power series expansion ${\log}(1-x)=-x-x^2/2-x^3/3-\cdots$ for
$x\in[-1,1]$ and the fact that, for $x\leq 1/2$,
\begin{eqnarray}
x+x^2/2+x^3/3+\cdots \leq x(1+x+x^2+\cdots) = x(1-x)^{-1} \leq 2x.
\end{eqnarray}
\end{example}



The cryptographic framework uses sets of type $A_{\delta}$, that
have, among others, the two properties defined in Eq. (\ref{pwdist})
and Eq. (\ref{nozerocomp}). The latter property, that
$\braket{0}{\psi}=0$ for all $\myket{\psi}$ in $A_{\delta}$, is
particular to our version of the framework. The former property of
pairwise $\delta$-almost-orthogonality is a reasonable condition to
impose on quantum-public-key cryptography in general, in order to
avoid a situation where two different private keys correspond to
practically indistinguishable physical scenarios. For example,
suppose Alice-$x$ has private key $x$ and Alice-$y$ has private key
$y$, and $x \neq y$; we would not want Alice-$y$ to be able to
convince Bob that she is Alice-$x$ with significant probability.
This property also rules out trivial lower-bound arguments based on
distinguishability, like in Example \ref{eg_BBSearch}, because an
$(A_{\delta},0)$-squasher \emph{does not increase pairwise
distinguishability}, by which we mean that, for all distinct
$\myket{\psi}$ and $\myket{\phi}$ in $A_{\delta}$,
\begin{eqnarray}\label{ineq_noincpwdist}
D(\sigma_{0,\psi},\sigma_{0,\phi}) \leq
D((\ketbra{\psi}{\psi})^{\otimes t},(\ketbra{\phi}{\phi})^{\otimes
t})
\end{eqnarray}
for some $t \in O(\textrm{polylog(M)})$; in the particular case of
$A_{\delta}$, the inequality (\ref{ineq_noincpwdist}) actually holds
for a constant $t$ (dependent on $\delta$). The set $A_2$ in Example
\ref{eg_BBSearch} has neither of the two properties discussed in
this paragraph, and we cannot see how to harness the provable
difficulty of squashing $A_2$ for potential cryptographic
application.


We now state the No-Squashing Conjecture more precisely.  Call
$A_{\delta}$ \emph{efficient} if and only if there is a succinct
description of $A_{\delta}$ (so that its classical description may
be easily published) and each state in $A_{\delta}$ is efficiently
computable given its classical description (i.e., although we do not
bound Eve's computational power, we prefer that Alice and Bob are
bounded by polynomial time). Since the trace distance between
$\sigma_{0,\psi}$ and the trivially-preparable mixture
$\ketbra{0}{0} + \ketbra{\psi}{\psi}$ is 1/2, we take $\epsilon<1/2$
in the conjecture.

\begin{conjecture}[No-Squashing]\label{conj_strongnosq}
There exists an efficient $A_{\delta}$ such that
\\$t(A_{\delta},\epsilon) \in \omega(({\log}(M))^k)$ for all
constants $k>0$ and any nonnegative constant $\epsilon < 1/2$.
\end{conjecture}

\noindent Our choice of the superpolynomial bound in the conjecture
corresponds to the usual meaning of ``large'' or ``inefficient'' in
complexity theory.  We do not claim that proving such a bound has
immediate cryptographic application: recall that no result about the
difficulty of squashing is sufficient for establishing the
honest-verifier security of the protocol of Section
\ref{sec_idscheme}.


%

We also leave as an open problem to find any set $A$ such that the
$(A,0)$-squasher does not increase pairwise distinguishability and
is not completely positive (or prove no such $A$ exists). Note that
the one-qubit universal-\textsc{not} gate \cite{BHW00}, which maps
\begin{eqnarray}
\alpha \myket{0} + \beta \myket{1} \mapsto \beta^* \myket{0} -
\alpha^* \myket{1}
\end{eqnarray}
for any qubit state $\alpha \myket{0} + \beta \myket{1}$, does not
increase pairwise distinguishability and is known to be not
completely positive.

\end{document}